\documentclass{PoS}
\usepackage{rotating}

\title{Complete calculation of evaluated Maxwellian-averaged cross sections and their uncertainties for s-process nucleosynthesis}

\ShortTitle{Complete calculation of evaluated Maxwellian-averaged cross sections and their uncertainties}

\author{\speaker{Boris Pritychenko}\\
        National Nuclear Data Center, Brookhaven National Laboratory, Upton, NY 11973-5000, USA\\
        E-mail: \email{pritychenko@bnl.gov}}

\abstract{Present contribution represents a significant improvement of our previous calculation of Maxwellian-averaged cross sections 
and astrophysical reaction rates.  Addition of newly-evaluated neutron reaction libraries, such as ROSFOND and Low-Fidelity Covariance 
Project, and improvements in  data processing  techniques allowed us to extend it for entire range of s-process nuclei, calculate 
Maxwellian-averaged  cross section uncertainties for the  first  time, and provide additional insights on all currently available 
neutron-induced reaction data.  Nuclear reaction calculations using ENDF libraries and current Java technologies will be discussed 
and new results will be presented.}

\FullConference{11th Symposium on Nuclei in the Cosmos, NIC XI\\
		July 19-23, 2010\\
		Heidelberg, Germany}

\begin{document}

\section{Introduction}
Nuclear reactions play an important role in stellar nucleosynthesis and are responsible for producing heavy chemical elements from 
light elements that were generated in the Big Bang. Present-day calculations of s-process nucleosynthesis are often based on  
dedicated nuclear astrophysics data tables, such as work of Bao et {\it al.} \cite{00Bao}, or its successor KADONIS \cite{06Dil}, however it is essential to produce complementary 
neutron-induced reaction data sets. ENDF-6 formatted evaluated neutron libraries contain various data for all known nuclei, 
including neutron capture cross sections for more than 680 individual nuclei from $^{1}$H to $^{257}$Fm in the range of neutron energy 
from 10$^{-5}$ eV to 20 MeV. Nuclear-reactor and national-security application communities used these data extensively in the 
eV and MeV energy ranges, while keV data were less utilized. This creates a unique opportunity to utilize  
evaluated neutron data for non-traditional intermediate-energy applications, such as s-process nucleosynthesis, and create a new set of ENDF benchmarks. This  work represents a significant upgrade of our previous calculation of Maxwellian-averaged cross sections  and 
 astrophysical reaction rates \cite{10Pri}.  In addition of newly-evaluated neutron reaction libraries, such as JENDL-4, CENDL-3.1, ROSFOND 2010 and Low-Fidelity Covariance Project \cite{07Zab,08Lit}, improvements in  data processing  techniques allowed us to extend calculations for the entire  range of s-process nuclei and produce Maxwellian-averaged  cross section uncertainties for the  first  time.
Nuclear reaction calculations using evaluated neutron libraries and current Java technologies will be discussed and new results will be presented.

\section{Calculation of Maxwellian-averaged Cross Sections and Uncertainties}
ENDF libraries are based on theoretical calculations that are often adjusted to fit experimental 
data \cite{exfor}. There are two kinds of ENDF cross section data representations: 
groupwise (averaged over energy interval) and pointwise. The first kind is often used in reactor physics 
calculations, while the second one is better suited for nuclear physics applications. Generic ENDF library 
cross sections (MF=3) do not contain information on neutron resonances. 
To resolve this problem for neutron physics calculations the codes PREPRO \cite{07Cul} and NJOY \cite{10Mac} are often used to produce a pointwise version of the libraries that include the resonance region data and provide cross 
section information within ENDF range of energies from 10$^{-5}$ eV to 20 MeV. Here, we used the code PREPRO to reconstruct the resonance region with a precision of 0.1$\%$.

Maxwellian-averaged cross sections can be expressed as \cite{10Pri}   

\begin{equation}
\label{myeq.max3}
\sigma^{Maxw}(kT) = \frac{2}{\sqrt{\pi}} \frac{(m_2/(m_1 + m_2))^{2}}{(kT)^{2}}  \int_{0}^{\infty} \sigma(E^{L}_{n})E^{L}_{n} e^{- \frac{E^{L}_{n} m_2}{kT(m_1 + m_2)}} dE^{L}_{n}
\end{equation}
where {\it k} and {\it T} are the Boltzmann constant and temperature of the system, respectively  and $E$ is an energy of 
relative motion of the neutron with respect to the target. Here  $E^{L}_{n}$ is a neutron energy in the laboratory system 
and $m_{1}$ and $m_{2}$ are masses of a neutron and target nucleus, respectively.

Previously \cite{10Pri}, Maxwellian-averaged cross sections and astrophysical reaction rates were produced using the Simpson method on linearized ENDF cross sections (MF=3). 
This simple method allowed quick calculated integral values with good precision. However the degree of precision was within $\sim$1$\%$ \cite{10Pri,05Nak}. 
This general limitation can be overcome in the linearized ENDF files because the cross section value is linearly-dependent on energy within a particular bin \cite{09Pri}
\begin{equation}
\label{myeq.int1}
\sigma (E) = \sigma_{1} + (E-E_1)\frac{\sigma (E_2) - \sigma (E_1)}{E_2 - E_1} 
\end{equation} 
where $\sigma (E_1),  E_1$ and  $\sigma (E_2),  E_2$ are pointwise cross section and energy values for the corresponding energy bin. Last equation is a good approximation of neutron cross section values for a sufficiently dense grid. This allowed us to calculate definite integrals using Wolfram Mathematica online integrator \cite{09Math}. Summing integrals for all energy bins will produce an exact integral value 
for Maxwellian-averaged cross section. Low-Fidelity cross section covariances were used to calculate  uncertainties for ENDF/B-VII.0 data \cite{08Lit,06Chad}. 
Final results for JENDL-4.0, ROSFOND 2010 and ENDF/B-VII.0 libraries \cite{Jendl,07Zab,06Chad} are shown in Table \ref{Table1}. 
Due to space limitations only selected data are shown and CENDL-3.1, JEFF-3.1 cross sections are omitted. 

\begin{table}
\centering
\caption[Evaluated nuclear libraries and KADONIS Maxwellian-averaged neutron capture cross sections.]{Evaluated nuclear libraries and KADONIS Maxwellian-averaged neutron capture cross sections in mb at {\it kT}=30 keV for  {\it s}-process nuclei.}
\label{Table1} 
\begin{tabular}{|c|cccc|}
\hline
Isotope &  JENDL-4.0  &   ROSFOND 2010  &  ENDF/B-VII.0    &  KADONIS \cite{06Dil} \\ 
\hline \hline
 36-Kr- 82 & 9.582E+1 &   9.483E+1 &  1.027E+2$\pm$2.097E+1   &  9.000E+1$\pm$6.000E+0 \\
 42-Mo- 96 & 1.052E+2 &   1.035E+2 &  1.036E+2$\pm$1.690E+1   &  1.120E+2$\pm$8.000E+0 \\
44-Ru-100 &  2.065E+2 &   2.062E+2 &  2.035E+2$\pm$3.949E+1   &  2.060E+2$\pm$1.300E+1 \\
46-Pd-104 &  2.700E+2 &   2.809E+2 &  2.809E+2$\pm$4.488E+1   &  2.890E+2$\pm$2.900E+1 \\
48-Cd-110 &  2.260E+2 &   2.346E+2 &  2.346E+2$\pm$4.219E+1   &  2.370E+2$\pm$2.000E+0 \\
50-Sn-116 &  9.115E+1 &   1.002E+2 &  1.002E+2$\pm$1.875E+1   &  9.160E+1$\pm$6.000E-1 \\
52-Te-122 &  2.644E+2 &   2.639E+2 &  2.349E+2$\pm$4.883E+1   &  2.950E+2$\pm$3.000E+0 \\
52-Te-123 &  8.138E+2 &   8.128E+2 &  8.063E+2$\pm$1.063E+2   &  8.320E+2$\pm$8.000E+0 \\
52-Te-124 &  1.474E+2 &   1.473E+2 &  1.351E+2$\pm$2.682E+1   &  1.550E+2$\pm$2.000E+0 \\
54-Xe-128 &  2.582E+2 &   2.826E+2 &  2.826E+2$\pm$6.823E+1   &  2.625E+2$\pm$3.700E+0 \\
54-Xe-130 &  1.333E+2 &   1.518E+2 &  1.518E+2$\pm$2.993E+1   &  1.320E+2$\pm$2.100E+0 \\
56-Ba-134 &  2.301E+2 &   2.270E+2 &  2.270E+2$\pm$4.038E+1   &  1.760E+2$\pm$5.600E+0 \\
56-Ba-136 &  7.071E+1 &   7.001E+1 &  7.001E+1$\pm$1.087E+1   &  6.120E+1$\pm$2.000E+0 \\
60-Nd-142 &  3.557E+1 &   3.701E+1 &  3.341E+1$\pm$4.252E+1   &  3.500E+1$\pm$7.000E-1 \\
62-Sm-148 &  2.361E+2 &   2.444E+2 &  2.449E+2$\pm$4.416E+1   &  2.410E+2$\pm$2.000E+0 \\
62-Sm-150 &  4.217E+2 &   4.079E+2 &  4.227E+2$\pm$3.601E+2   &  4.220E+2$\pm$4.000E+0 \\
64-Gd-154 &  9.926E+2 &   1.010E+3 &  9.511E+2$\pm$1.070E+2   &  1.028E+3$\pm$1.200E+1 \\
66-Dy-160 &  8.702E+2 &   8.293E+2 &  8.328E+2$\pm$6.769E+1   &  8.900E+2$\pm$1.200E+1 \\
72-Hf-176 &  5.930E+2 &   4.529E+2 &  4.571E+2$\pm$4.811E+1   &  6.260E+2$\pm$1.100E+1 \\
80-Hg-198 &  1.612E+2 &   1.612E+2 &  1.612E+2$\pm$1.621E+1   &  1.730E+2$\pm$1.500E+1 \\
82-Pb-204 &  8.355E+1 &   7.242E+1 &  7.242E+1$\pm$7.699E+0   &  8.100E+1$\pm$2.300E+0 \\
\hline
\end{tabular}

\end{table}

In {\it s-}process nucleosynthesis, we assume that product of neutron-capture cross section (at 30 keV in mb) times solar system abundances (relative to Si = 10$^6$) as a function of atomic mass should be constant for equilibrium nuclei \cite{88Rol}: 
\begin{equation} 
\label{myeq.eq} 
\sigma_{A}N_{A}= \sigma_{A-1}N_{A-1} = constant
\end{equation}
 
\begin{figure}
\begin{center}
\includegraphics[height=7cm]{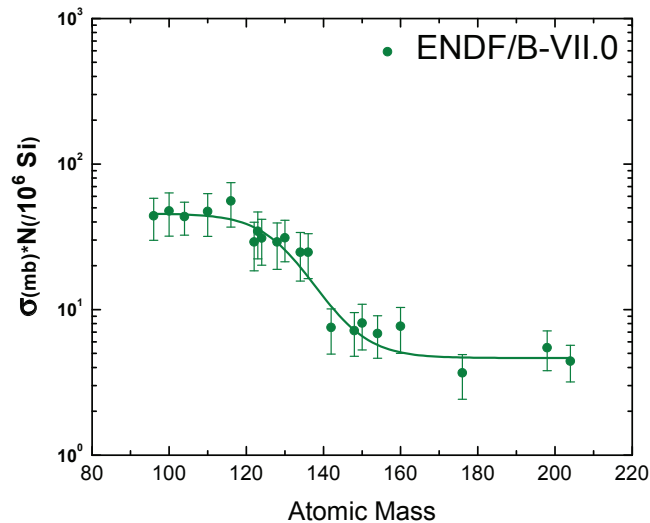}
\caption{ENDF/B-VII.0 product of neutron-capture cross section (at 30 keV in mb) times solar system abundances (relative to Si = 10$^6$) as a function of atomic mass for nuclei produced only in the {\it s}-process.}
\label{fig1}
\end{center}
\end{figure}

To verify this phenomenon, the calculated $\langle \sigma^{Maxw}_{\gamma} (30 keV) \rangle$ from the ENDF/B-VII.0 library \cite{06Chad} were multiplied by solar abundances taken from 
 Anders and  Grevesse \cite{89And}, and plotted  in Figure \ref{fig1}. 
Visual inspection of the Figure indicates two local equilibrium and ledge-precipice break at $A \sim$ 138 for the ENDF/B-VII.0 fit. 
Relatively high product value for $^{116}$Sn is due to the fact that $^{116}$Sn has {\it r-}process contribution \cite{89And}.

\begin{figure}
\begin{center}
\includegraphics[height=12cm]{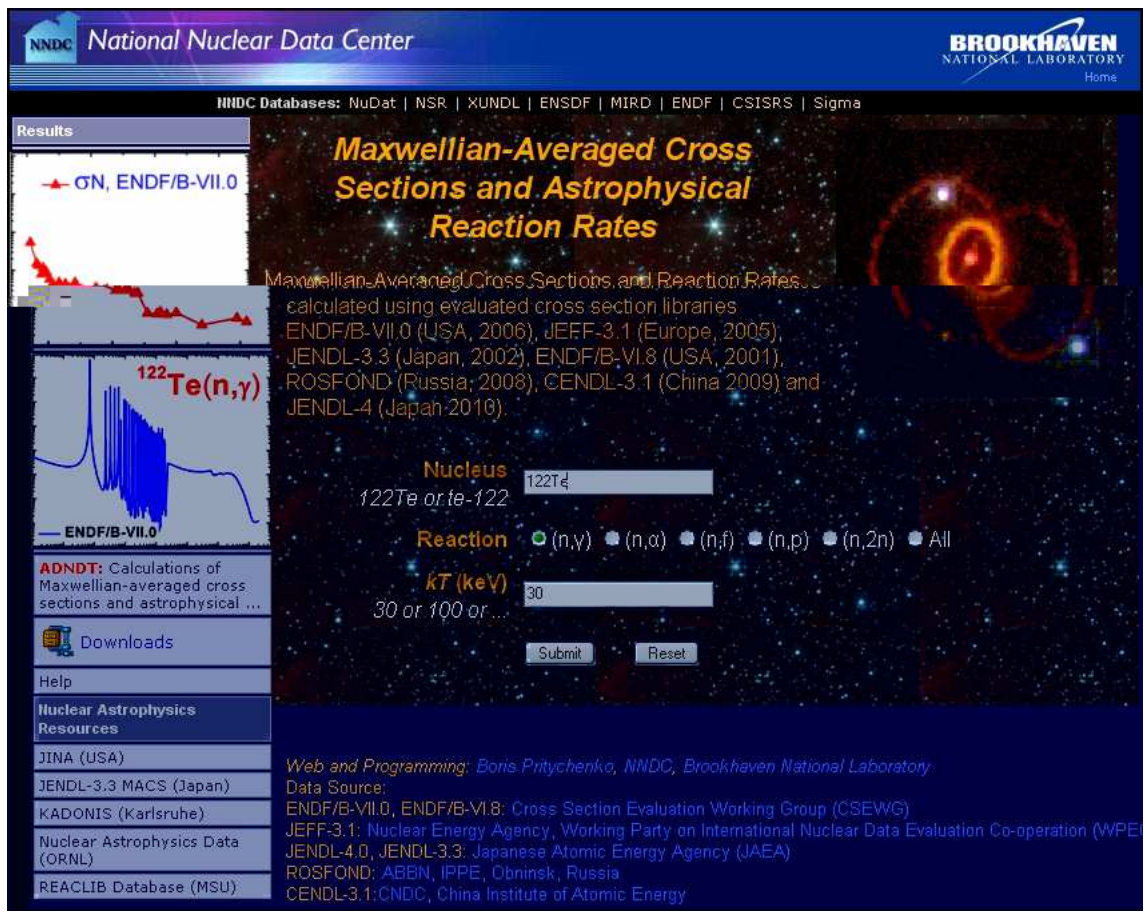}
\caption{`Maxwellian-averaged Cross Sections and Astrophysical Reaction Rates' Web application, {\it http://www.nndc.bnl.gov/astro}.}
\label{fig2}
\end{center}
\end{figure}

In FY 2011, present contribution results will be used to upgrade `Maxwellian-averaged Cross Sections and Astrophysical Reaction Rates' Web application {\it http://www.nndc.bnl.gov/astro}. The Web application frontpage is shown in Figure \ref{fig2}.

\section{Conclusion \& Outlook}
Maxwellian-averaged cross sections for neutron capture have been calculated using JENDL-4, CENDL-3.1, ROSFOND 2010, ENDF/B-VII.0 and JEFF-3.1 libraries using Low-Fidelity covariance project data.  ROSFOND 2010 library (686 materials) \cite{07Zab} provides 
the most complete coverage along the s-process path. In fact, it contains information on 347 out of 354 isotopes (98\% coverage). The only {\it s-}process isotopes 
that are missing in ROSFOND are: $^{12}$C,  $^{110}$Ag, $^{132,133,135,137}$Ce and $^{142}$Pr. Carbon and Praseodymium cross sections can be found in the CENDL-3.1 and ENDF/B-VII.0 or 
JEFF-3.1 libraries, respectively. Typical neutron library, such as ENDF/B-VII.0 (393 materials), provides ~80\% coverage of 
the {\it s-}process path.

The comparison of Maxwellian-averaged cross section values from Table \ref{Table1} indicates a good agreement between KADONIS  and 
evaluated nuclear libraries.  Future work on ENDF libraries will provide additional improvements and
benefits for nuclear astrophysics and applications communities.

\section{Acknowledgments}
The author thanks Drs. V. Zerkin (IAEA), M.W. Herman,  C.M. Mattoon, S.F. Mughabghab, and M. Pigni (BNL) for productive discussions and help on ROSFOND and Low-Fidelity project data, respectively.  We are also grateful to M. Blennau for  a careful reading of the manuscript.
This work was sponsored by the Office of Nuclear Physics, Office of Science of the U.S. Department of Energy, under Contract 
No. DE-AC02-98CH10886 with Brookhaven Science Associates, LLC.

\end{document}